\documentclass[journal,draftcls,onecolumn,12pt,twoside]{IEEEtranTCOM}

\usepackage{epsfig}
\usepackage{graphicx}
\usepackage{epstopdf}
\usepackage{amsmath}
\usepackage{amsfonts}
\usepackage{amssymb}
\usepackage{setspace}
\usepackage{pgf,pgfarrows,pgfnodes,pgfautomata,pgfheaps,pgfshade}
\usepackage{tikz}
\usepackage{amsthm}
\usepackage[english]{babel}
\usepackage{algorithm}
\usepackage{algpseudocode}
\usepackage{bbm}
\usepackage{nomencl}
\makenomenclature
\theoremstyle{plain}
\newtheorem{thm}{Theorem}[section]

\algdef{SE}[DOWHILE]{Do}{doWhile}{\algorithmicdo}[1]{\algorithmicwhile\ #1}%
\makeatletter
\def\BState{\State\hskip-\ALG@thistlm}
\makeatother
\renewcommand{\thesection}{\arabic{section}}
\allowdisplaybreaks
\begin{document}
\title{Resource Allocation in a MAC with and without security via Game Theoretic Learning}

\author{Shahid M Shah,~\IEEEmembership{Student~Member,~IEEE,}, Krishna Chaitanya A
	and~Vinod~Sharma,~\IEEEmembership{Senior~Member,~IEEE}
\thanks{Part of the paper was presented in IEEE Information theory and applications (ITA) workshop 2016, La Jolla, San Diego, USA}
\thanks{Shahid M Shah Krishna Chaitanya A, and Vinod Sharma are with Electrical communication Department, Indian Institute of Science, Bangalore, India.}}
\maketitle

\begin{abstract}
	In this paper we study a $K$-user fading MAC, with and without an eavesdropper (Eve). In the system without Eve, we assume that each user knows only its own channel gain and is completely ignorant about the other users' channel state. The legitimate receiver sends a short acknowledgement message ACK if the message is correctly decoded and a NACK if the message is not correctly decoded. Under these assumptions we use game theoretic learning setup to make transmitters \textit{learn} about the power allocation under each state. We use multiplicative weight \textit{no-regret} algorithm to achieve an $\epsilon$-coarse correlated equilibrium. 
	 We also consider the case where a user can receive other users' ACK/NACK messages. Now we can maximize a weighted sum-utility and achieve Pareto optimal points. 
	 We also obtain Nash bargaining solutions, which are Pareto points that are fairer to the transmitting users. 
	
	 With Eve, we first assume each user knows only its own channel gain to the receiver as well as to Eve. The receiver decides whether to send an ACK or a NACK to the transmitting user based on the secrecy-rate condition. We use the above developed algorithms to get the equilibrium points.
	 Next we study the case where  each user knows only the \textit{distribution} of the channel state of Eve.
\end{abstract}


\vspace{0.01cm}
\allowdisplaybreaks
\section{Introduction}
A multiple access channel (MAC) is a basic building block in wireless networks \cite{cover2012elements}. Also, it models the uplink in a wireless cellular system. Therefore it has been studied extensively over the years (\cite{ahlswede1973multi}, \cite{liao1972multiple}, \cite{el2011network}). More recent, it has also received attention from information theoretic security point of view. In this paper, we study a MAC with and without an eavesdropper using game theoretic techniques. This allows operating in the capacity region which is fair to the users and also provides distributed algorithms with local information at the users. First we provide a literature survey on this problem.

 A general $M-$user fading MAC is considered in \cite{tse1998multiaccess} where the receiver has perfect channel state knowledge and broadcasts channel state information of all the users to all the transmitters. The authors prove that the capacity region of a $M-$user MAC has a polymatroid structure, and they exploit this structural property to find the optimal power and rate control policy.  Time varying additive white Gaussian noise (AWGN) MAC is studied in \cite{shamai1997information} where it is assumed that only the receiver can track the channel and not the transmitters. In that case the transmitters allocate fixed powers (which satisfy the average power constraint) and transmit data over the channel.
 
 In \cite{hesham2008water} the authors propose a distributed power allocation scheme using \textit{Game Theory}. The authors assume that each user knows the channel gain of other users also, in addition to knowing his own channel gain. The authors prove that the sum-rate point on the capacity region is a Nash equilibrium when the decoding strategy of the receiver is not known to the transmitters. The authors also prove the existence of a Stackelberg equilibrium in which the receiver acts as a leader and the transmitters play a low level game. Using repeated games, the authors prove that each point on the capacity region of a fading MAC is achieved by some power control policy.  In \cite{altman2010bayesian} the authors prove stronger results by assuming that each user knows only its own channel gain, but knows the distribution of channel gains of the other users. Under these conditions, the authors prove the existence and uniqueness of a \textit{Bayesian equilibrium}. In an orthogonal multiple access channel the authors in \cite{learningmac2012distributed} have used evolutionary game theory to obtain a power allocation scheme, while assuming that each user knows the channel gain of all users via feedback.
\par
With security constraints, a multiple access wiretap channel (MAC-WT) has been well studied in literature. One of the early works is reported in \cite{yates2006} where only one user has confidential messages to be transmitted. The authors have obtained upper bounds on the secrecy-rate regions. In \cite{liang2008multiple} the authors consider a more general setup wherein they consider a discrete memoryless multiple access channel where the transmitting users receive a noisy version of each others' conversation, and they do not trust each other. In this scenerio the authors have obtained an achievable secrecy rate region and some outer bounds. In some special cases this provides secrecy capacity region. A multiple access wiretap channel with feedback has been studied in \cite{tang2007multiple}. An achievable region of a Gaussian multiple access wiretap channel (G-MAC-WT) was obtained in \cite{tekin2008gaussian} (the secrecy capacity region is still an open problem). 
  \par
  In the above work, weak secrecy criterion is used. A strong secrecy based achievable rate region for a MAC-WT is reported in \cite{yassaee2010multiple}. In \cite{bagherikaram2010secure} the authors find secure degrees of freedom for a MAC-WT. More recently in \cite{zivari2014compound} the authors have studied a compound MAC-WT and have characterized inner and outer bounds on the secrecy capacity region. In \cite{shah2012achievable} the authors have studied a fading MAC-WT with full CSI of Eve and also when each user knows the channel state of all the users to the receiver, but is ignorant of the instantaneous value of channel state to the eavesdropper (only its distribution is known). But knowing other users' channel gains to the legitimate receiver may also not be practical: it needs a lot of signalling overhead and feedback information. Hence in this paper we present a game-theoretic solution to the resource allocation scheme under the hypothesis that each user only knows its own channel gain and is completely ignorant of other users' channels (not even the distributions).
 \par
 In interference channel model \cite{chaitanyalearning2015learning}, the authors use learning algorithms to study a stochastic game, and learn optimal power allocation policies. The authors use no-regret algorithm to prove the existence of a \textit{correlated equilibrium}. It is assumed that each user knows power allocation policy of other users, which is not always realistic. The same authors extend this work to the case where each user knows only his own channel gain and does not know the power levels used by other users. The authors prove the existence of a coarse correlated equilibrium using \textit{multiplicative weights} no-regret algorithm (\cite{chaitanya2015distributed})
 
In this paper we first consider a fading MAC (F-MAC) without security constraint. We assume each user knows only its individual channel gain (unlike \cite{altman2010bayesian} we do not assume that it knows the distributions of channel gains of others). Since the receiver is receiving data from all the users, it is quite practical to assume that the receiver has channel state information of all the transmitting users. Once a user sends a codeword corresponding to a particular message, the receiver sends an ACK if it decodes it successfully, else it sends a NACK. Each user defines a utility based on the ACK/NACK. We use multiplicative weight no-regret algorithm to obtain an equilibrium. We also assume in the later part of the paper, that each user can decode ACK/NACK of other users and hence knows their utility. Then we aim to maximize the sum-utility and propose an algorithm to obtain a Pareto point. We also find a Nash bargaining solution which provides a Pareto point and ensures fairness among users. We also study the case where users can transmit at multiple rates rather than fixed rates.

Next we consider a fading MAC-WT where we first assume that each user knows its channel gains to the receiver and Eve. In this case we repeat all the algorithms which we used for a F-MAC (without security), i.e., MW, PP, NBS and also consider the multiple rates case.
 Since it is not practical to assume instantaneous channel gain of the eavesdropper to be known at the transmitter and the receiver, we next consider the case where the receiver only knows the distribution of the Eve's channel gains. The receiver calculates secrecy-outage and sends an ACK/NACK based on that. We again obtain a CCE, PP and a NBS.
 To the best of our knowledge this is the first paper which is using game theory on MAC-WT.
  Finally we compare the sum-rates obtained via all these algorithms to the global CSI case and also with the sum-rate obtained in \cite{shah2012achievable}.
  
 The rest of the paper is organized as follows. In Section 5.2 we describe the channel model and formulate the problem. In Section 5.3 we use Multiplicative Weight Algorithm to obtain a CCE. In Section 5.4 we obtain Pareto optimal points. In Section 5.5 we a consider fading-MAC-WT when the CSI of Eve is not available at the transmitters (only its distribution is known) and obtain a CCE, a NBS, and a PP. In Section 5.6 we compare the various schemes on an example. Finally, in Section 5.7 we conclude the paper.

\section{Fading MAC: Without Security Constraint}
 A time slotted F-MAC channel is considered with $K$ users who have messages to be transmitted to a receiver. Let $\{\widetilde{H}_i(t)\}$ be the channel gain process from user $i$ to the receiver at time $t$. User $i$ transmits $X_i(t)$ and the receiver receives 
 \begin{align}
 Y(t)&=\sum_{i=1}^{K}\widetilde{H}_i(t)X_i(t)+\eta_b(t),
 \end{align}
at time $t$, where $\eta_b(t)$ is white Gaussian noise with mean zero and variance 1, denoted by $\mathcal{N}(0,1)$, and independent of $\{X_i(t)\}$ and $\{\tilde{H}_i(t)\}$. Let $H_i(t)\triangleq \lvert \widetilde{H}_i(t) \rvert^2$. The fading gains are assumed discrete valued, in the sets $\mathcal{H}_i \triangleq \{h_i^{(1)}, \ldots, h_i^{(M)}\}$. Also $\{H_i(t),t\geq 0\}$ are independent and identically distributed ($iid$) sequences with distributions $\{\alpha_i^{(1)}, \ldots, \alpha_i^{(M)}\}$. To transmit any codeword, user $i$ can choose any power level from the set $\mathcal{P}_i \triangleq \{P_i^{(1)},\ldots, P_i^{(M)}\}$. Also, user $i$ has average power constraint $\overline{P}_i$.
\par
 User $i$  transmits at a fixed rate $r_i$ (to be generalized later) via a usual point to point channel encoder. If the receiver successfully decodes a message, it sends an ACK to that particular user. Otherwise, it sends a NACK. We assume that the NACK, ACK are transmitted at low rates so that these can be received with negligible error at the intended transmitter. The goal of each user is to maximize its probability of successful transmission.  
  \par
  Each user $i$ is assumed to know its own channel gain $H_i(t)$  at time $t$. Since the receiver can estimate the channel gain of all the users (either by receiving known pilots or by using initial data received), the receiver can use successive cancellation decoding strategy to decode all the users.
  \par
 Let $\pi(i)$ be the user which has the $i^{th}$ highest channel gain (in case of a tie we arbitrarily order them).
   The decoder first decodes the user $\pi(1)$ with the best channel gain first, taking the transmissions from the other users as noise. Then it removes it from the received signal $Y(t)$ and then decodes the next best user, taking the other users as noise and so on. 
   Let
  \begin{align}
  C_b\left(P_{\pi(i)},P_{-\pi(i)},H_{\pi(i)}\right)\triangleq \frac{1}{2}\log\left(1+\frac{H_{\pi(i)}P_{\pi(i)}(H_{\pi(i)})}{1+\sum_{j=i+1}^K H_{\pi(j)}P_{\pi(j)}(H_{\pi(j)})}\right).
  \end{align}
  Then the receiver will send an ACK to the transmitting user $\pi(i)$ if
  \begin{align}
  r_{\pi(i)} &\leq C_b(P_{\pi(i)},h_{\pi(i)},g_{\pi(i)}).
  \end{align}
  The above constraint follows from the successive cancellation decoding scheme chosen.
 Each user $i$ takes action (allocating power) $P_i^{(j)}$ when its channel gain is $H_i^{(j)}$ to transmit at its rate. We define feasible action space for user $i$ as
 \begin{align}
 \mathcal{P}_i= \left\lbrace \mathbf{P}_i=(P_i^{(1)},\ldots,P_i^{(M)}): P_i^{(k)}\in \{p_i^{(1)},\ldots,p_i^{(M)}\}, \sum_{j=1}^M \alpha_i(j)P_i^{(j)}\leq \overline{P}_i \right\rbrace.
 \end{align}
 We define $\lvert \mathcal{P}_i \rvert \triangleq M_i$ (where $\lvert A \rvert$ denotes the cardinality of set $A$) and index the elements of set $\mathcal{P}_i$ as $\{1,\ldots,M_i\}$. Let $a_i$ denote a feasible power policy of user $i$, i.e., $a_i$ takes a value from $\mathcal{P}_i$, and $a_i(h)$ is the power level used by user $i$ when its channel gain is $h\in \mathcal{H}$ under policy $a_i$. The action space of $K$ users is denoted as 
 
\begin{equation}
 \mathcal{P}=\bigotimes_{i=1}^{K}\mathcal{P}_{i},
\end{equation} and the action space of users, other than user $i$ is 
\begin{equation}
 \mathcal{P}_{-i}=\bigotimes_{j=1, j\neq i}^{K}\mathcal{P}_{j},
\end{equation} 
 where $\bigotimes_{i=1}^{N}A_i=A_1 \times A_2  \ldots \times A_N$. 
 The action profile of all the users is denoted as $a=(a_1,\ldots,a_K)$. A probability distribution $\psi(i)$ on $\mathcal{P}_{i}$ is called a strategy of user $i$. When a certain action is chosen with probability one, it is called a \textit{pure strategy}. 
 The objective of each transmitter is to maximize its probability of successful transmission. Since the actions chosen by one user may influence the outcome for the other users in terms of probability of successful transmission, this can be formulated as a stochastic game. For user $i$, if the channel gain in time slot $t$ is $H_{i}(t)$ and the action profile chosen is $(a_{i},a_{-i})$, we define its reward as,
 \[\omega_{i}^{(t)}\left(a_{i}^{(t)},H_{i}(t)\right)=
 \begin{cases}
 1,~\text{if user}~i~\text{receives an ACK}, \\
 0,~\text{otherwise}.
 \end{cases}
 \]
 We are interested in the time average of the reward process 
 \begin{equation}
 \nu _{i}\left( a_{i},a_{-i} \right)= \limsup_{T\rightarrow \infty}\frac{1}{T}\sum_{t=1}^T \omega_{i}^{(t)}\left(a_{i},H_{i}(t)\right).
 \end{equation}
 We will restrict ourselves to Markov stationary policies, i.e., action of user $i$ depends only on its current state $H_i(t)$. Then $\{\omega_{i}(a_{i},H_{i}(t))\}$ are $iid$ across time $t$. Hence by strong law of large numbers, the average reward $\nu_{i}(a_{i},a_{-i})=\mathrm{E}\left[\omega_{i}^{(t)}\left(a_{i},H_{i} \right)\right]$ is same as the probability of successful transmission.
 In terms of a mixed strategy $(\psi_{i},\psi_{-i})$, the average reward is 
 \begin{align}
 \nu _{i}&\left(\psi_{i},\psi_{-i}\right) =\sum_{a\in \mathcal{P}}\left[\prod_{j=1}^{K} \psi_{\pi(j)}\left(a_{\pi(j)}\right)\right]\nu _{i}\left(a_{i},a_{-i}\right). \label{utility_new_F_MAC_WT}
 \end{align}
 \par
 Hence this stochastic game can be modelled as a one-shot game in which player $i$ maximizes its utility (\ref{utility_new_F_MAC_WT}). In the rest of the paper we develop algorithms to compute equilibrium points for this game.
 \section{Multiplicative Weight Algorithm for Learning CCE} 
 In this section we use multiplicative weight algorithm (\cite{arora2012multiplicative}) to compute an equilibrium point of the system. This is a distributed algorithm. The cost of each user can be defined as $C_{i}((a_{i},a_{-i})\triangleq -\nu _{i}(a_{i},a_{-i})$. Now we have the following definition.
 
 \textbf{Definition 2}: If a distribution $\psi$ on $\mathcal{P}$ satisfies 
 \begin{equation}
 \mathsf{E}_{a \sim \psi}\left[C_{i}\left(a\right)\right]\leq \mathsf{E}_{a \sim \psi}\left[C_{i}\left(\hat{a}_{i},a_{-i}\right)\right]  +\epsilon,
 \end{equation}
 for each $i$ and all actions $\hat{a}_i$, then it is called $\epsilon-$ coarse correlated equilibrium, where on the right side $a_{-i}$ has the marginal distribution $\psi$.
 
 A mixed-Nash equilibrium is a CCE. Hence for our finite game a CCE exists (\cite{arora2012multiplicative}).
 
 \textbf{Definition 3:}(\cite{arora2012multiplicative}) For user $i$, the external regret is defined as 
 \begin{align}
 \frac{1}{T}\sum_{t=1}^T \mathsf{E}_{a_{-i\sim \psi_{-i}}} \left[C_{i}^{(t)}\left(a_{i}^{(t)},a_{-i}\right)  -C_{i}^{(t)}\left(a_{i},a_{-i}\right) \right]
 \end{align}
 for a given pure strategy sequence $a_{i}^{(1)},\ldots,a_{i}^{(T)}$ with respect to an action $a_{i}$.
 
 In a \textit{No-regret} algorithm, called multiplicative weight algorithm, users update their strategies based on the cost received, such that the external regret coverges to zero. This algorithm is presented in Algorithm 1. It converges to a CCE according to the following theorem  (\cite{arora2012multiplicative}, \cite{cesa2006prediction}).
 \begin{algorithm}
 \caption{Multiplicative Weights Algorithm}\label{euclid}
 \begin{algorithmic}[1]
 \Do
 \Procedure{Weight Update}{}
 \State $w^{(t)}_{i}(a_i) \gets 1,~\forall~a_i,~i=1,2,\ldots,K$
 \State User $i$: Choose action $w.p.$ $\Phi_i^{(t)}=\frac{\omega_i^{(t)}(a_i)}{\sum_{\hat{a}\in\mathcal{P}_i}w_i^{(t)}(\hat{a}_i)}$
 \BState Time $t$
 \State User $i$ receives average utility for choosing $a_i$ $\nu_i^{(t)}=\mathsf{E}_{a_{-i}\sim \Phi_{-i}}[C(a_i,a_{-1})]$
 \BState Update the weight
 \State $\quad$ $w_i^{(t+1)}(a_i)=w_i^{(t)}(a_i)(1-\epsilon)^{c_i^{(t)}(a_i)}$
 \BState Time $t+1$
 \State $\quad$ Calculate $\psi_t=\prod_{i=1}^K \Phi^{(t)}_i$
 \EndProcedure
 \doWhile{$\frac{1}{T}\sum_{t=1}^T \mathsf{E}_{a_{-i}\sim \psi_{-i}} \left[C_{i}^{(t)}(a_{i}^{(t)},a_{-i})  -C_{i}^{(t)}\left(a_{i},a_{-i}\right) \right] \linebreak >\epsilon$}
 \end{algorithmic}
 \end{algorithm}
  
 \begin{thm}
 Let $\psi^{(t)}=\prod_{i=1}^K \Phi_{i}^{(t)}$ denote the outcome distribution at time $t$. There exists an integer $T>0$ such that the regret of user $i$ is less than $\epsilon$ after $T$ iterations. Then, $\psi =\frac{1}{T}\sum_{t=1}^{T} \psi^{(t)}$ is an $\epsilon$-coarse correlated equilibrium.
 \end{thm}
 
 \section{Pareto Optimal Points}
 In a wireless environment it is realistic to assume that the ACK/NACK bits sent to a particular user can be successfully decoded by all the other users also (because these are sent at a low rate using robust codes). In that case all users can learn about the utility of each other at time $t$. We show in this section that this information can be used to get a socially optimal Pareto point which generally provides a better performance than a CCE.
 
 \textbf{Definition}: An action profile $a\in \mathcal{P}$ is a \textit{Pareto point} if there does not exist another profile $\tilde{a}$ such that $\nu_i(\tilde{a})\geq \nu_i(a)$, $\forall~i\in \mathcal{K}$ and $\nu_j(\tilde{a})>\nu_i(a)$ for some $j\neq i$.
 \par
  Define
 \begin{equation}
 \Omega(a)=\sum_{i=1}^K \gamma_i \nu_i(a),
 \end{equation}
 for fixed $\gamma_i \geq 0,~i=1,\ldots,K$. Then a solution to the optimization problem
 \begin{align}
 \max_{a}~~ \Omega(a), \nonumber ~~ \text{subject to}~&a\in \mathcal{P}. \label{pareto_max}
 \end{align}
 is a Pareto point (\cite{miettinen2012nonlinear}).

 In Algorithm 2 below we provide a distributed algorithm in which the users update their strategies in a sequential fashion so as to improve $\Omega(a)$.
 This distributed algorithm is the variation of a heuristic stochastic local search algorithm. In this algorithm each user chooses a random action and uses it for a fixed number of time slots (say $T$). Then each user finds weighted sum of the utilities (since each user receives ACK/NACK of other users, it can calculate this quantity). After $T$ slots a user experiments randomly (with probability say, $\rho$) and then with some probability updates the action profile according to its channel state. Now one user uses this action for next $T$ slots and the other users use the previous action. Based on the weighted sum of utilities, the particular user defines a benchmark. The details of algorithm in the scenario of interference channel can be found in \cite{chaitanyalearning2015learning}. The algorithm is presented below as Algorithm 2.

 \begin{algorithm}
 \caption{Distributed Algorithm to obtain Pareto Points}\label{euclid1}
 \begin{algorithmic}[1]
 \State User i: choose $a_i\in \mathcal{P}_i$ uniformly.
 \State Use $a_i$ for $T$ time slots.
 \Procedure{Weight Update}{}
 \State Update weight of each user $i$
 \State $\hat{\Omega}(a)\gets \sum_{i=1}^K \gamma_i \left (\frac{1}{T} \sum_{t=1}^T \omega_i^{(t)}(a_i,H_i(t),G_i(t))\right)$
 \State After $T$ slots: $w.p$ $\rho_i$ user $i$ experiments
 \Procedure{Action Update}{}
 \State $w.p$ $\epsilon$ choose $a_i^{\prime}\neq a_i$, $a_i^{\prime}\in \mathcal{P}_i$
 \State $w.p.$ $1-\epsilon$
 \State \quad choose $a_i^{\prime} \neq a_i$ s.t. $h_i$ with high $\alpha_i$ gets higher power level
 \State If $\alpha_i$ same for all $h_i$, then higher value of channel state gets higher power level.
 \EndProcedure
  \State Call new action $\hat{a}_i$
  \State User $i$: use $\hat{a}_i$ for $T$ time slots.
  \State $\hat{a}_j=a_j$ if user $j$ is not experimenting.
  \State User $i$: find $\hat{\Omega}(\hat{a}_i,a_{-i})$
  \State \If{$\hat{\Omega}(\hat{a}_i,\hat{a}_{-i})>\hat{\Omega}(a_i,\hat{a}_{-i})$}
  $a_i \gets \hat{a}_i$
  \State $P_{benchmark}=\hat{\Omega}(\hat{a}_i,\hat{a}_{-i})$
  \Else 
  \State Randomly select another action
  \EndIf
  \EndProcedure
 \end{algorithmic}
 \end{algorithm}
 \section{Nash Bargaining Solution}
 The Pareto points obtained in Section 4 are socially optimal, but may not be fair to all users: some users may get much more rates than others. To obtain fair Pareto points we use the concept of Nash Bargaining Solution (NBS) \cite{nash1950bargaining}.
 \par
  In NBS we need to specify a \textit{disagreement} strategy $\Delta$ and the corresponding outcome $\delta=(\delta_1,\ldots,\delta_K)$ that specifies the utility of each user that it receives by playing the disagreement strategy whenever there is no improvement over this utility in playing the bargaining outcome. We define the set of all possible utilities as
 \begin{equation}
 \mathcal{V}=\{(\nu_1(a),\ldots,\nu_K(a)): a\in \mathcal{P}\}.
 \end{equation}
 This bargaining problem is denoted by $(\mathcal{V},\delta)$. 
 \par
 The aim of the bargaining problem is to find a bargaining solution which is Pareto optimal and satisfies the axioms of symmetry, invariance and independence of irrelevant alternatives (\cite{boche2009nash}).
 \begin{thm}[\cite{nash1950bargaining}]
 There exists a unique bargaining solution (provided the feasible region is non-empty) and it is given by the solution of the optimization problem:
 \begin{align}
 &\max ~\prod_{i=1}^{K}(\nu_i-\delta_i)\nonumber \\ 
 &\mathsf{subject~ to}~ \nu_i\geq \delta_i, i=1,\ldots,K ,~(\nu_1,\ldots,\nu_{K})\in\mathcal{V}. \label{NASH_MAX} \square
 \end{align}
 \end{thm}

  We obtain the disagreement outcome for our problem by the following procedure
  
\begin{itemize}
\item[$\star$] Each user chooses an action that gives higher power level to the channel state that has higher probability
 of occurrence. In other words, among the set of feasible actions, choose a subset of pure strategies that
 gives the highest power level to the channel state with highest probability of occurrence. We shrink
 the subset by considering the actions that give higher power level to the second frequently occurring
 channel state and we repeat this process until we get a single strategy.
 \item[$\star$] If all the channel states occur with equal probability, we follow the above procedure by considering
 the value of the channel gain instead of the probabilities of occurrence of the channel gains.
\end{itemize}

  Let $a_i$ denote the pure strategy chosen by the $i^{th}$ user and let $T_{\delta}$ be the number of time slots over which this strategy is used. Then the disagreement value for user $i$ is 
  \begin{equation}
  \delta_i=\frac{1}{T_{\delta}}\sum_{t=1}^{T_{\delta}}\omega_i^{(t)}(a_i,H_i(t)).
  \end{equation}

 We use Algorithm 2 to obtain a distributed solution of (\ref{NASH_MAX}), with the objective function defined as
 \begin{equation}
 \Omega(a)=\prod_{i=1}^{K}\left(\nu_i(a)-\delta_i\right).
 \end{equation}
 From \cite{nash1950bargaining}, if the set of utilities $\mathcal{V}$ is convex then a Nash bargaining solution is also \textit{proportionally fair}. In our problem $\mathcal{V}$ is convex and hence the solution is proportionally fair also.

 \section{Fading MAC With security constraints}
 In this section we consider a time slotted fading-MAC-WT channel with $K$-users who have messages to  transmit confidentially to a legitimate receiver (Bob), while a passive eavesdropper (Eve) is listening to the conversation and trying to decode. The notation corresponding to Bob is same as in the previous sections. Here we define the notation for the channel to Eve. Let $\{\widetilde{G}_i(t)\}$ be the channel gain process from user $i$ to Eve. At time $t$ Eve receives  
  \begin{align}
  Z(t)=\sum_{i=1}^{K}\widetilde{G}_i(t)X_i(t)+\eta_e(t),
  \end{align}
  where $\eta_e(t)$ is white Gaussian noise, with distribution  $\mathcal{N}(0,1)$ and independent of $\{\eta_b(t)\}$ and the channel gain processes and $\{X_i(t)\}$. We define $G_i(t)\triangleq \lvert \widetilde{G}_i(t) \rvert^2$. The fading gains of Eves' channels are assumed discrete valued, in the set  $\mathcal{G}_i \triangleq \{g_i^{(1)},\ldots,g_i^{(M)}\}$. Also   $\{G_i(t),t\geq 0\}$ are $iid$ independent of each other and also of the sequences $\{H_i(t)\}$, with distribution   $\{\beta_i^{(1)},\ldots,\beta_i^{(M)}\}$ respectively.   User $i$  transmits at a fixed rate $r_i$ via wiretap coding. If the receiver successfully decodes (see details below in this subsection), it sends an (ACK) to that particular user. Otherwise it sends a NACK. We assume that the NACK, ACK are transmitted at low rates so that these can be received with negligible error at the intended transmitter. The goal of each user is to maximize the probability of successful transmission.  
  \par
  Each user $i$ is assumed to know its own channel gains $H_i(t)$ and $G_i(t)$ at time $t$. Since the receiver can estimate the channel gain of all the users (either by receiving known pilots or by using initial data received), the receiver can use successive decoding strategy to decode all the users.

   We define
  \begin{align}
  \hspace*{-30pt}
    C_e\left(P_{\pi(i)},P_{-\pi(i)},H_{\pi(i)},G_{\pi(i)}\right)\triangleq \frac{1}{2}\log\left(1+\frac{G_{\pi(i)}P_{\pi(i)}(H_{\pi(i)},G_{\pi(i)})}{1+\sum_{j\neq i}^K G_{\pi(j)}P_{\pi(j)}(H_{\pi(j)},G_{\pi(j)})}\right)
  \end{align}
  Then the receiver will send an ACK to the transmitting user $\pi(i)$ if
  \begin{align}
  \hspace*{-20pt}
  r_{\pi(i)} &\leq \left(C_b(P_{\pi(i)},h_{\pi(i)},g_{\pi(i)})-C_e(P_{\pi(i)},h_{\pi(i)},g_{\pi(i)})\right)^+,\nonumber \\
  \end{align}
  otherwise a NACK, where $(a)^+=\max(0,a)$.
  The above constraint follows from the achievable secrecy-rate region of a Gaussian MAC-WT as discussed in \cite{tekin2008gaussian}.
 Each user $i$ takes action (allocating power) $P_i^{(j)}$ when its channel gains are $H_i^{(j)}$ and $G_i^{(j)}$ to transmit at its rate.

 Now we can use all the algorithms of Section II to obtain a CCE, PP and NBS.
 \subsection{Fading MAC-WT with Individual Main Channel CSI Only}
 We consider now the case where the users as well as the receiver do not know Eve's channel gain, but only its distribution. Also the transmitters \textit{do not know even the distribution} of Eve's channel gains. In this scenario, the natural metric for the receiver to decide whether to send an ACK or a NACK will be outage based. First we define the secrecy outage, when $h_1,\ldots,h_K$ are given, as

 \begin{align}
 \hspace{-50pt}
 & P_O^S(\pi(i))\triangleq \nonumber \\
 & Pr\left\{ r_\pi(i)> \log\left( 1+ \frac{h_{\pi(i)}P_{\pi(i)}(H_{\pi(i)})}{1+\sum_{j=i+1}^K h_{\pi(j)}P_{\pi(j)}(H_{\pi(i)})}\right)   -\log\left( 1+ \frac{G_{\pi(i)}P_{\pi(i)}(H_{\pi(i)})}{1+\sum_{j\neq i}^K G_{\pi(j)}P_{\pi(j)}(H_{\pi(j)})}\right)\right\}.
 \end{align}
 The receiver sends an ACK if $P_o^S<\epsilon$, else the receiver sends a NACK. Hence we define utility of user $i$ as
 \begin{equation}
 \omega_{i}\left(a_{i}^{(t)},h_{i}(t) \right)=\mathbbm{1}_{\{P_O^S(i)<\epsilon\}}
 \end{equation}
 where $\mathbbm{1}_{\{C\}}$ is an indicator function. With these utility functions, we can use the algorithms provided in Sections III-V.

  \subsection{Avoiding Security Breach}
  In the previous sections we assumed that when the legitimate receiver cannot securely decode the message it sends a NACK. This is useful for the transmitters to learn the equilibrium point. But the messages transmitted during those slots may be decoded by Eve (with probability $> \epsilon$ in Section 6A). Now we modify the system a little so as to use the above coding scheme but mitigate this secrecy loss also.
  \par
  We assume that each slot is comprised of two subslots. The fading process does not change during the whole slot. In the first part of the slot we transmit a dummy (random) message. If Bob sends an ACK to user $i$ then the actual confidential message can be transmitted by user $i$ in the second subslot at the same power. If Bob sends a NACK then user $i$ should not use the second subslot. We can make the second subslot much larger than the first subslot so that the rate loss due to the dummy messages is minimal.
 
 \section{Transmission at Multiple Rates}
 Till now we have considered the case where the users are transmitting at fixed rates. Now we consider the more realistic scenario where the users can transmit at different rates, depending on their channel gains. We assume that user $i$ can choose any rate from the rate set $\mathcal{R}_i=\{r_i^{(1)},\ldots, r_i^{(M_R)}\}$. We now define a new strategy set such that choosing the rate of transmission becomes part of the action taken along with the power chosen. Hence we define the modified strategy set as
 \begin{align}
 \mathcal{A}_i\triangleq &\left\lbrace (r_i,P_i^{(1)},\ldots,P_i^{(M)}):  r_i \in \mathcal{R}_i, P_i^{(k)}\in \{p_i^{(1)},\ldots,p_i^{(M)}\},  \sum_{j=1}^M \alpha_i(j)\beta_i^{(j)}P_i^{(j)}\leq \overline{P}_i \right\rbrace
 \end{align}
 We can now use all the existing algorithms to compute CCE, PP and NBS.

\section{Numerical results}

In this section we provide several examples using the algorithms developed in this paper.
We divide our examples into two parts: 1) F-MAC (without security constraint) and 2) F-MAC-WT.
\subsubsection*{F-MAC (without security constraint)}
\begin{figure}
   \hspace{-30pt}
   \includegraphics[scale=0.54]{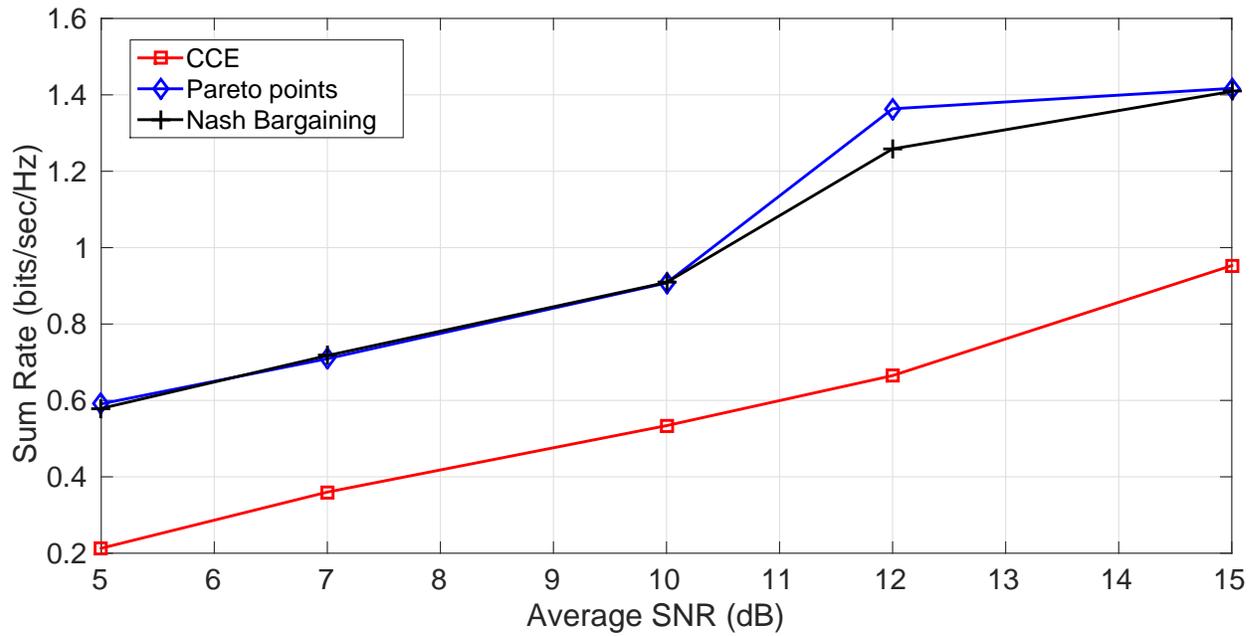}
   \caption{Sum-rate comparison: CCE vs NBS vs PP (F-MAC, fixed rate case).}
   \label{F_MAC_FR_COMPARE}
 \end{figure}

\begin{figure}
   \hspace{-25pt}
     \includegraphics[scale=0.55]{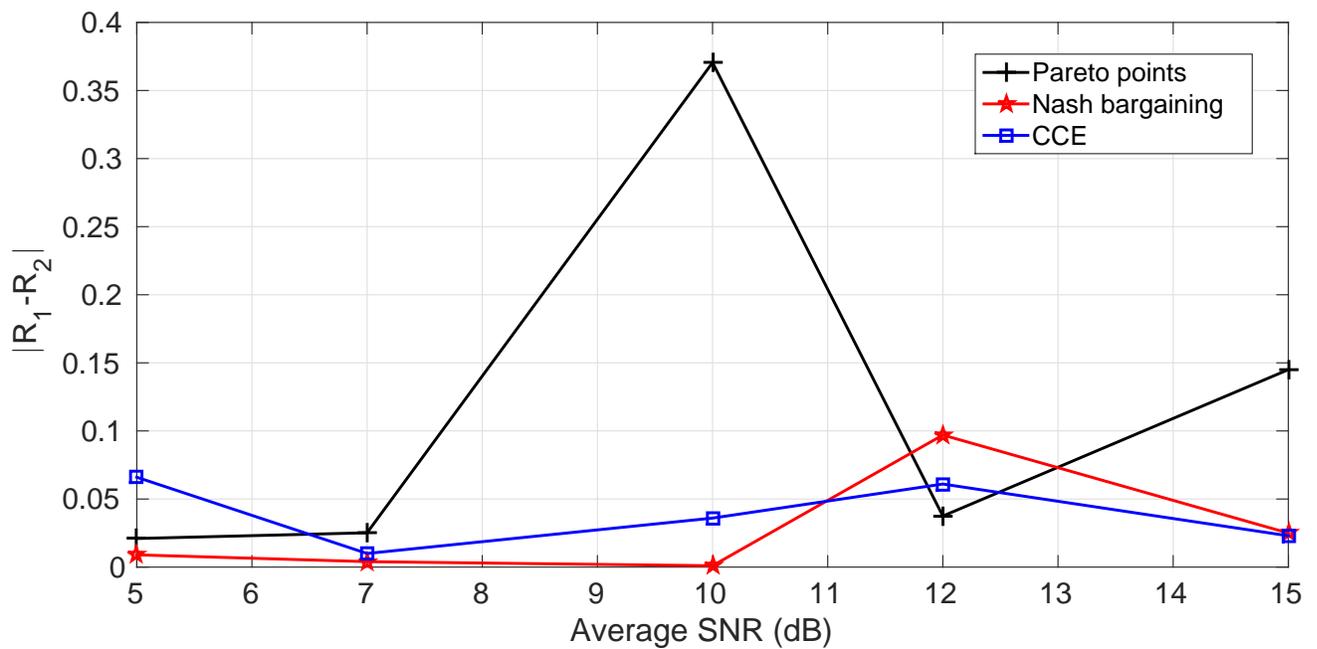}
    \caption{Fairness comparison for FMAC fixed rate.}
    \label{F_MAC_FR_fairness}
    \end{figure}

We first consider a fading MAC where we take $\mathcal{H}=\{0.1,0.5,0.9\}$ chosen with uniform distribution over the set, for all users and we assume that a user can choose any power from the power set $\{1,5,\ldots,100\}$. In this scenario we first consider the case when users are transmitting at fixed rate, 1 bit/sec. In this scenario we compare the sum-rate obtained by our three algorithms i.e., CCE, PP and NBS (see Fig. \ref{F_MAC_FR_COMPARE}). We note that NBS and PP are better than CCE. Also, regarding the fairness among the users, we see from Fig. \ref{F_MAC_FR_fairness} that NBS is fairer than PP and CCE.
\par
\begin{figure}
   \hspace{-25pt}
     \includegraphics[scale=0.55]{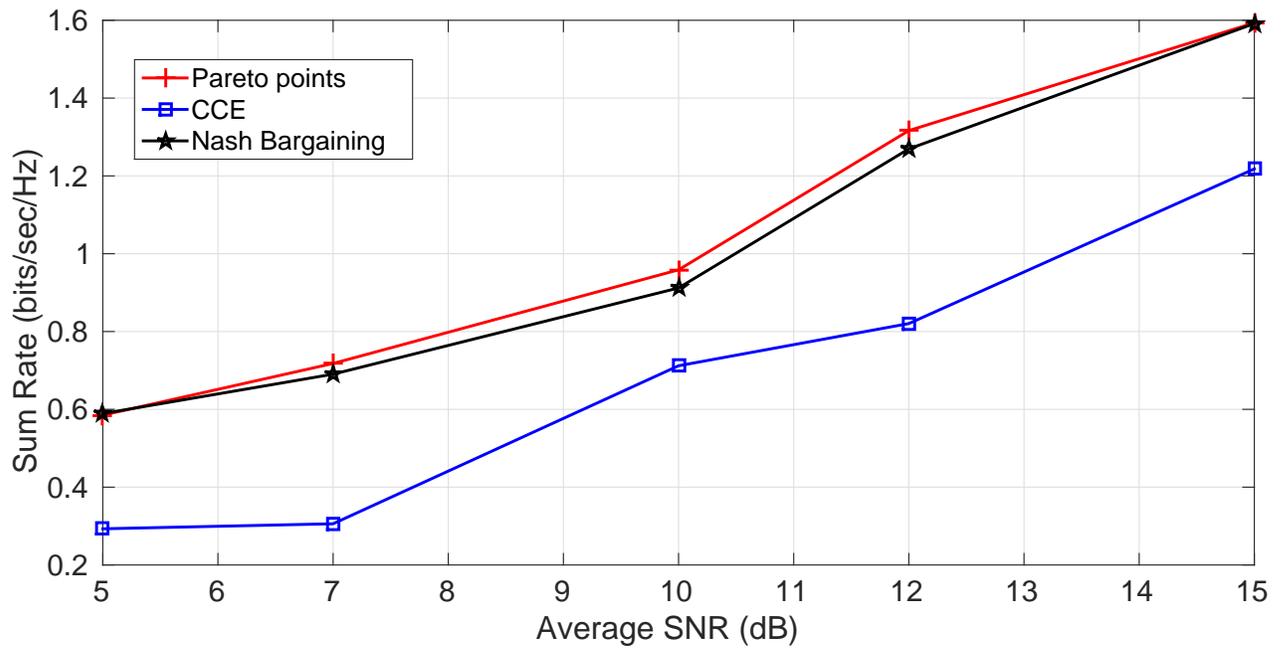}
      \caption{Sum-rate comparison for FMAC: multiple transmission rates.}
      \label{F_MAC_MR_compare}
      \end{figure}
\begin{figure}
   \hspace{-35pt}
     \includegraphics[scale=0.55]{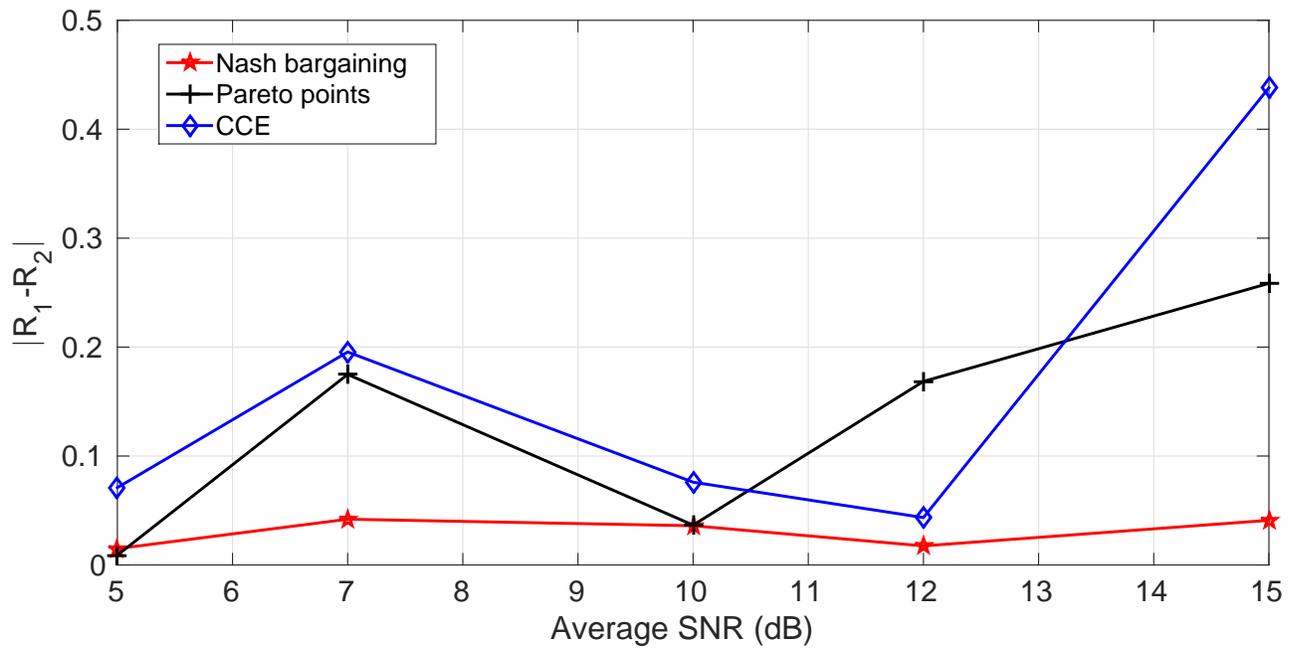}
      \caption{Fairness comparison for F-MAC: multiple transmission rates.}
      \label{F_MAC_MR_fairness}
      \end{figure}

Next we consider a more practical case where users can choose transmission rates from the set $\{0.4,~ 0.8,~ 1,~ 1.5,~ 2,~ 2.3\}$. Here also we compare the sum-rate obtained via CCE, PP and NBS. To get the result for CCE all users use Algorithm 1. For finding Pareto points, all users use Algorithm 2, with the weights $\gamma_i=1$. As expected, we observe that PP and NBS give much better rates than CCE (Fig. \ref{F_MAC_MR_compare}). From Fig. \ref{F_MAC_MR_fairness} we also observe that here also NBS is fairer among the three algorithms.
\par
\begin{figure}
   \hspace{-25pt}
          \includegraphics[scale=0.55]{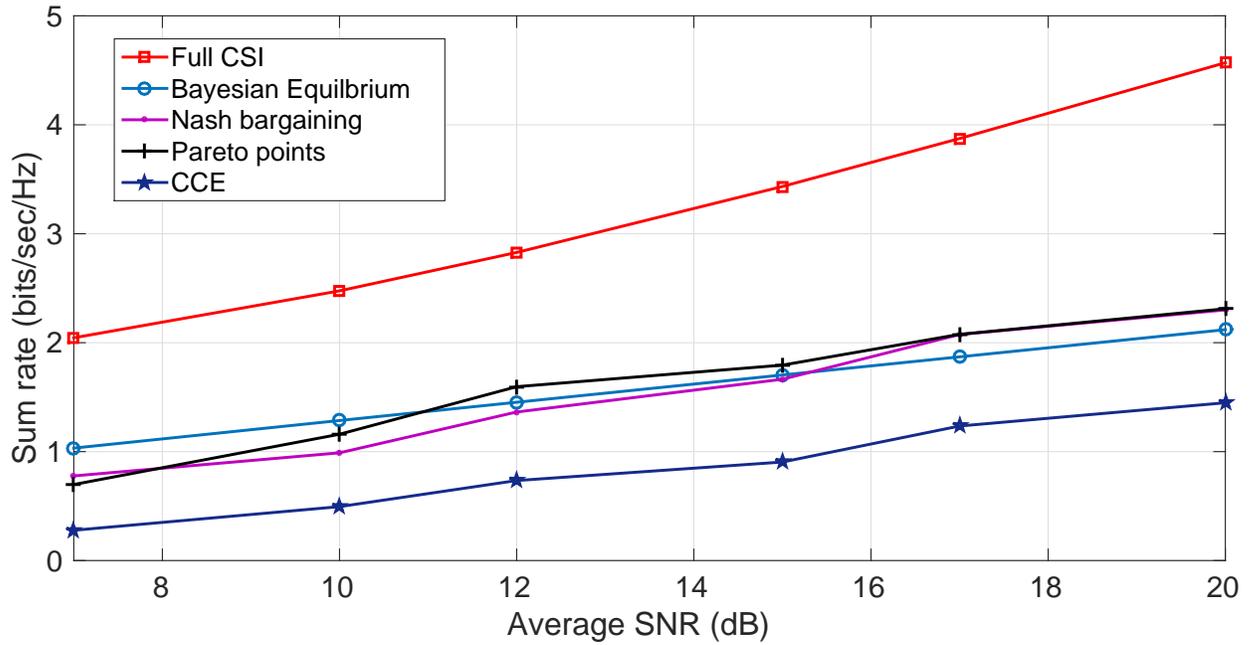}
          \caption{F-MAC: Sum-rate comparison for our scheme vs existing schemes.}
          \label{F_MAC_VS_ALTMAN}
          \end{figure}

Finally, to compare the performance with the existing schemes, we take an example where we assume $\mathcal{H}=\{0.1,~0.9\}$ and the power set is $\{1,5,\ldots,100\}$. Also as in the previous example, the users can choose transmission rates from the set $\{0.4,~ 0.8,~ 1,~ 1.5,~ 2,~ 2.3\}$. We compare our algorithms (viz. CCE, PP and NBS) with the case where global knowledge of CSI is assumed. We also compare our schemes with that of \cite{altman2010bayesian}, where each user knows its own channel and distribution of other users' channel gains. We observe that PP and NBS give better sum-rate than this scheme (Fig. \ref{F_MAC_VS_ALTMAN}).

\subsubsection*{F-MAC-WT (with security constraint)}
\begin{figure}
   \hspace{-25pt}
     \includegraphics[scale=0.55]{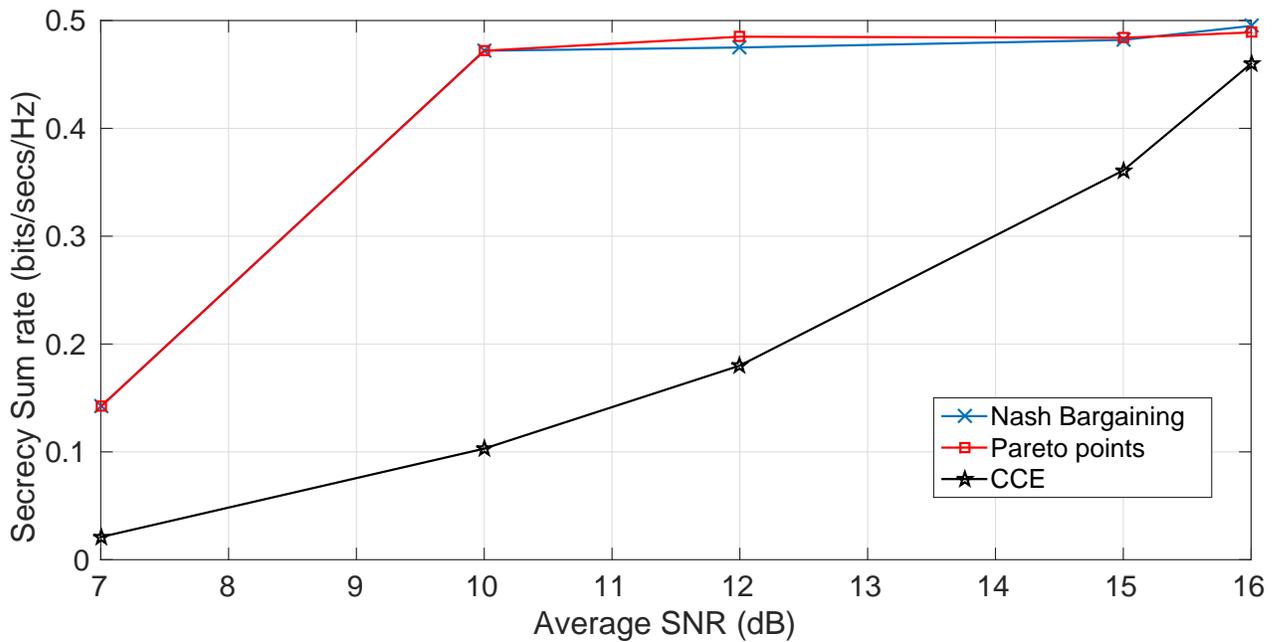}
           \caption{Sum-rate with security constraints: comparison of CCE, PP and NBS at fixed transmission rate (with CSI of Eve).}
           \label{F_MAC_WT_FR_FCSI_COMP}
           \vspace{-0.0 cm}
           \end{figure}

\begin{figure}
   \hspace{-25pt}
     \includegraphics[scale=0.55]{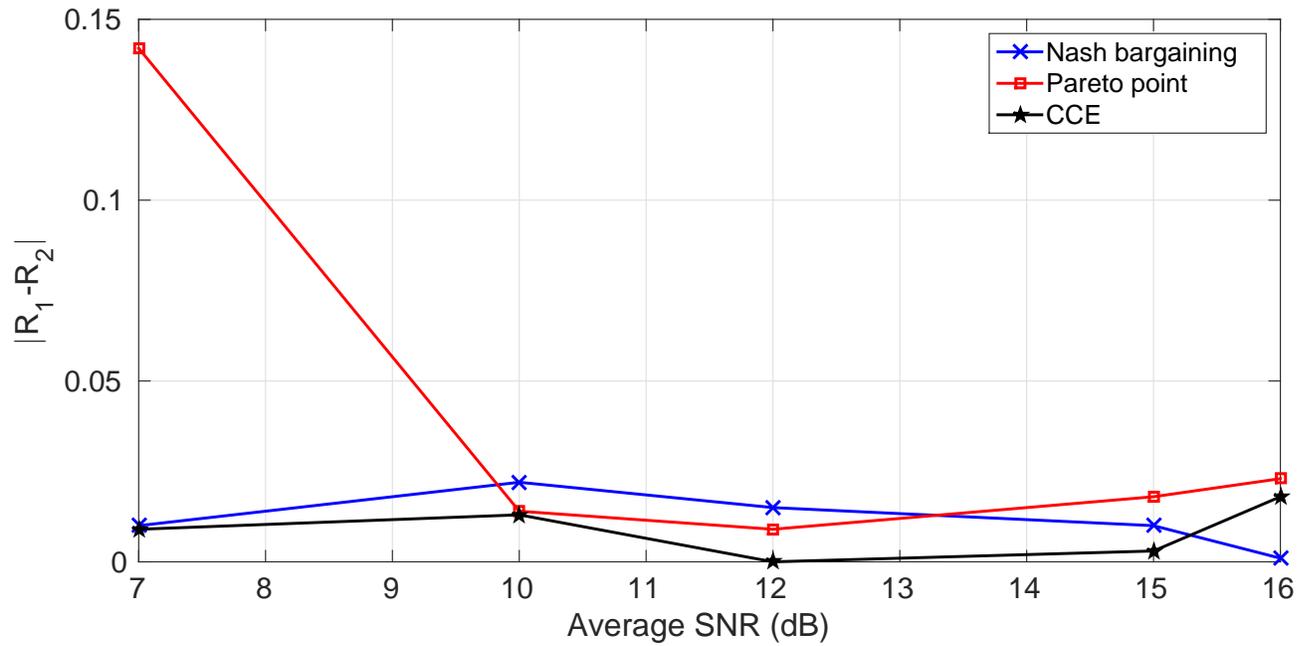}
            \caption{Fairness of CCE, PP and NBS at fixed transmission rate (with CSI of Eve)}
            \label{F_MAC_WT_FR_FAIRNESS_FCSI}
            \vspace{-0.0 cm}
            \end{figure}

Next we  consider a 2-user fading MAC-WT with full CSI. We let $\mathcal{H}=\{0.1,0.5,0.9\}$ and $\mathcal{G}=\{0.05,0.4,0.8\}$ for both the users. We assume that the probability with which any state can occur is equal, i.e., $\alpha_i^j=\beta_i^j=1/3$ for $i=1,2$, and $j=1,2,3$. A user can choose any power from the power set $\{1,5,\ldots,100\}$. 
We first consider a fixed rate scenario. Each user knows its channel gain to Bob and Eve. We observe that the PP and the NBS obtain much higher sum rate than the CCE (Fig. \ref{F_MAC_WT_FR_FCSI_COMP}). Also we observe that the NBS is fairer than the PP and the CCE (Fig. \ref{F_MAC_WT_FR_FAIRNESS_FCSI}). 
\par
\begin{figure}
     		\hspace{-25pt}
     		  \includegraphics[scale=0.55]{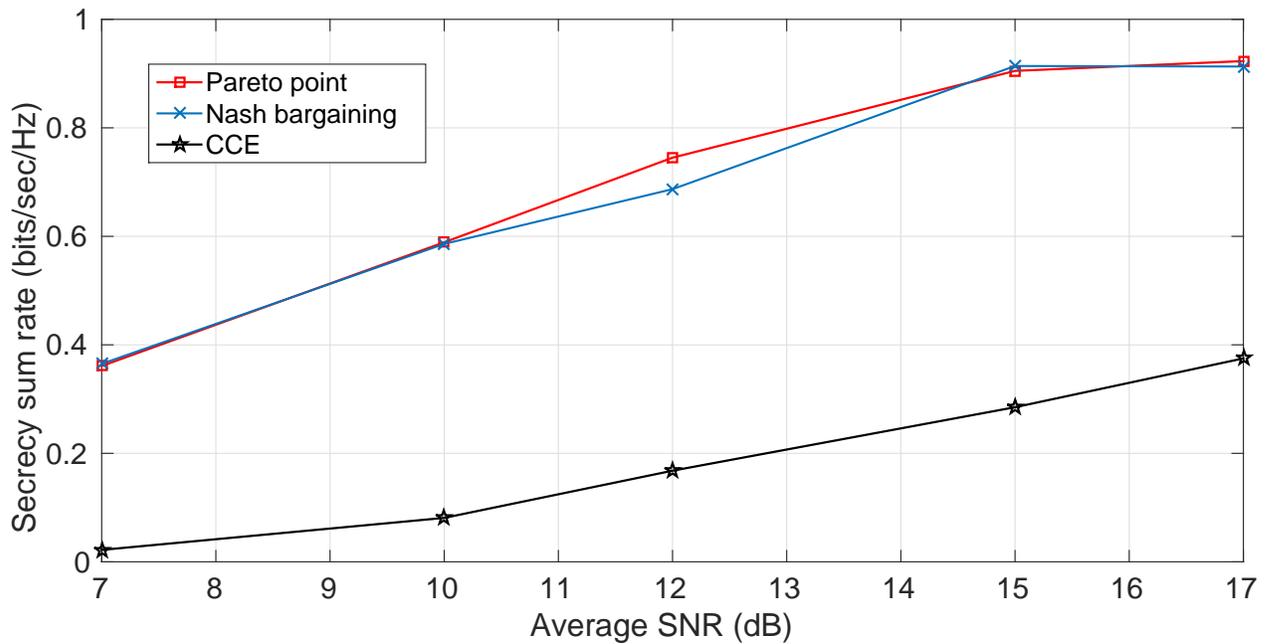}
              \caption{comparison of CCE, PP and NBS for F-MAC-WT, with no CSI of Eve (Fixed transmission rate)}
              \label{F_MAC_WT_FR_COMPARE_NOCSI}
              \end{figure}

Next we consider the case where the users don't have CSI of Eve available but only the distribution is known. As in the previous example, here also we observe the same trend (Fig. \ref{F_MAC_WT_FR_COMPARE_NOCSI}, Fig. \ref{F_MAC_WT_FR_FAIRNESS_NOCSI}).
\par
\begin{figure}
   \hspace{-25pt}
     \includegraphics[scale=0.55]{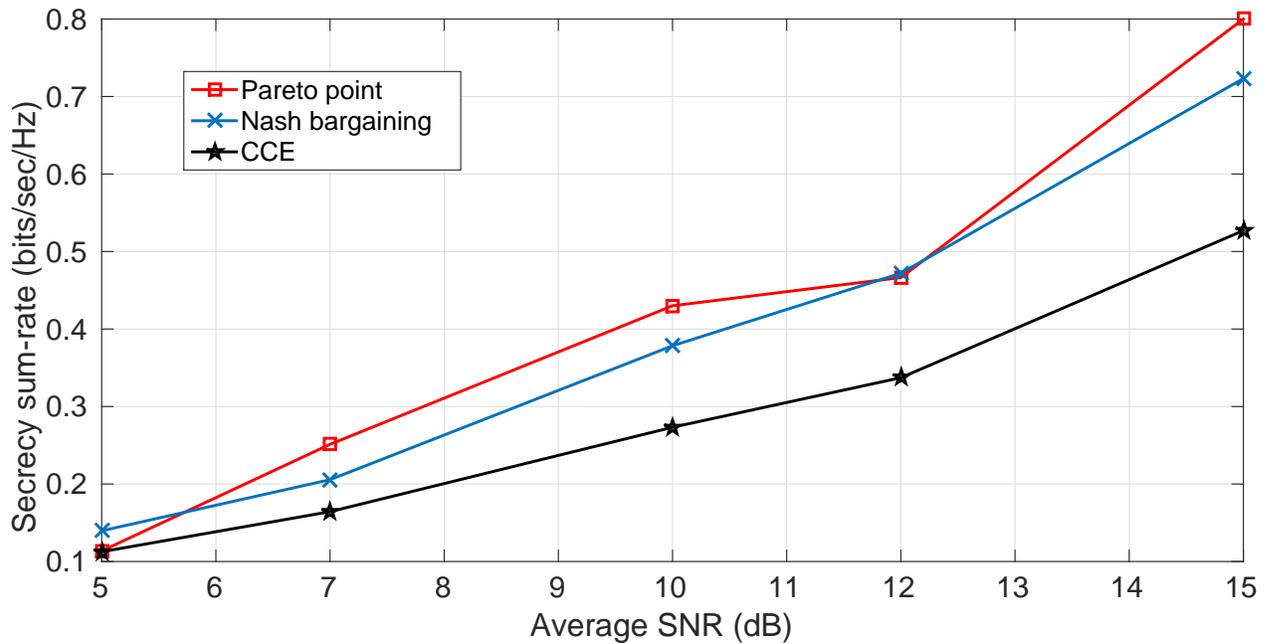}
             \caption{FMAC-WT: Sum-rate comparison of CCE, PP and NBS for multiple rate case (with CSI of Eve)}
             \label{F_MAC_WT_MR_COMP_FCSI}
             \vspace{-0.0 cm}
     \end{figure}
\begin{figure}
   \hspace{-25pt}
     \includegraphics[scale=0.55]{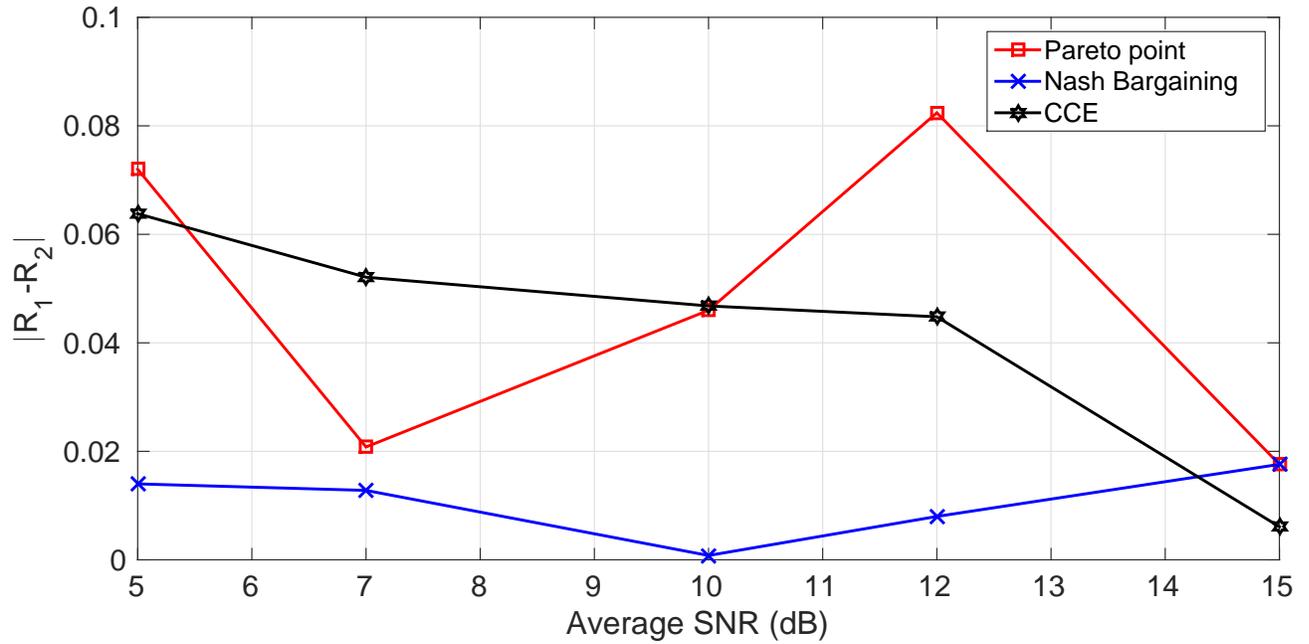}
             \caption{F-MAC-WT: Fairness comparison of CCE, PP and NBS for multiple rate case (with CSI of Eve).}
             \label{F_MAC_WT_MR_FAIRNESS_FCSI}
             \vspace{-0.0 cm}
             \end{figure}

Next we consider the case when users have CSI of Eve available to them and can transmit at multiple rates choosing from $\{0.1,~ 0.2,~ 0.3,~ 0.4,~ 0.5,~ 0.6\}$. From Fig. \ref{F_MAC_WT_MR_COMP_FCSI} we note that PP and NBS give better secrecy sum-rates and from Fig. \ref{F_MAC_WT_MR_FAIRNESS_FCSI} we observe fairness of NBS.
\par
\begin{figure}
   		\hspace{-25pt}
            \includegraphics[scale=0.55]{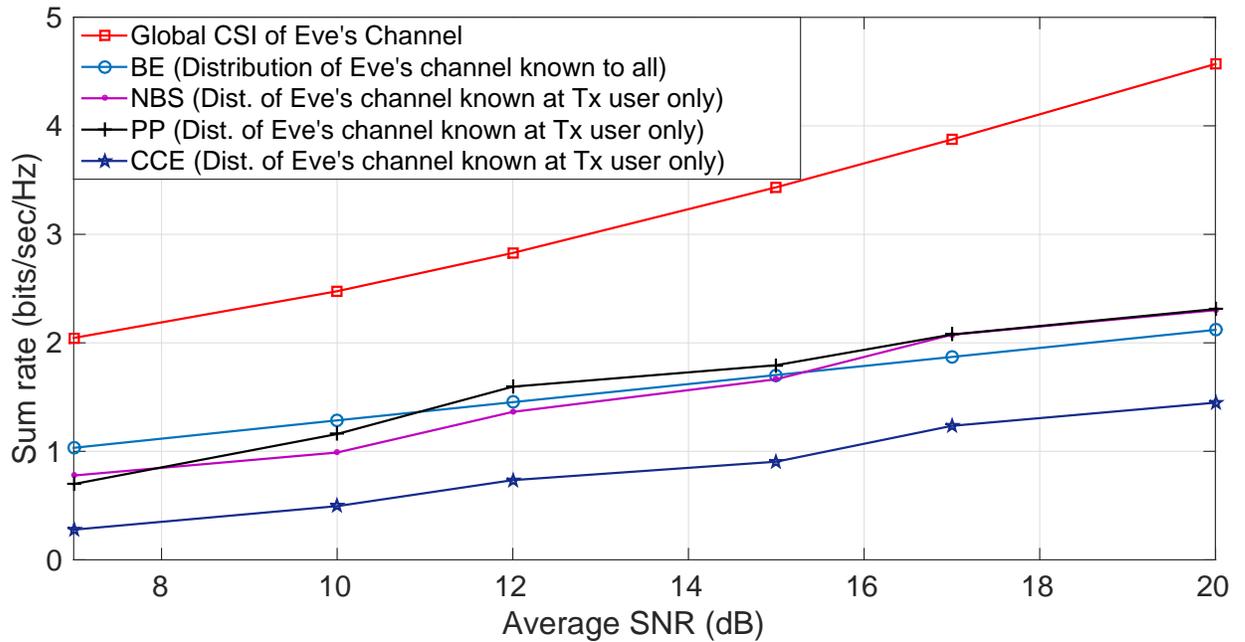}
            \caption{F-MAC-WT: Comparing with existing schemes}
            \label{F_MAC_WT_EXISTING}
            \end{figure}

 
We take one more example with $\mathcal{H}=\{0.1,.9\}$ and $\mathcal{G}=\{0.05,0.8\}$. We compare the NBS and the PP with the case when CSI of the transmitters is known globally but only the distribution of Eve's channel gains are known at all transmitters. This case is studied in \cite{shah2012achievable} for continuous channel states and a centralized solution which maximizes the sum rate is found. We also find the Bayesian Equilibrium (BE) for the case when each user knows distribution of all the channel gains to Eve, as done in \cite{altman2010bayesian} for F-MAC without security constraints. Here we observe that the NBS and the PP outperform BE  at high SNR (Fig. \ref{F_MAC_WT_EXISTING}). At low SNR the sum-rate for the NBS and the PP are quite close to that of BE. We also observe here that the CCE performs the worst.


     \begin{figure}
      		\hspace{-25pt}
      		  \includegraphics[scale=0.55]{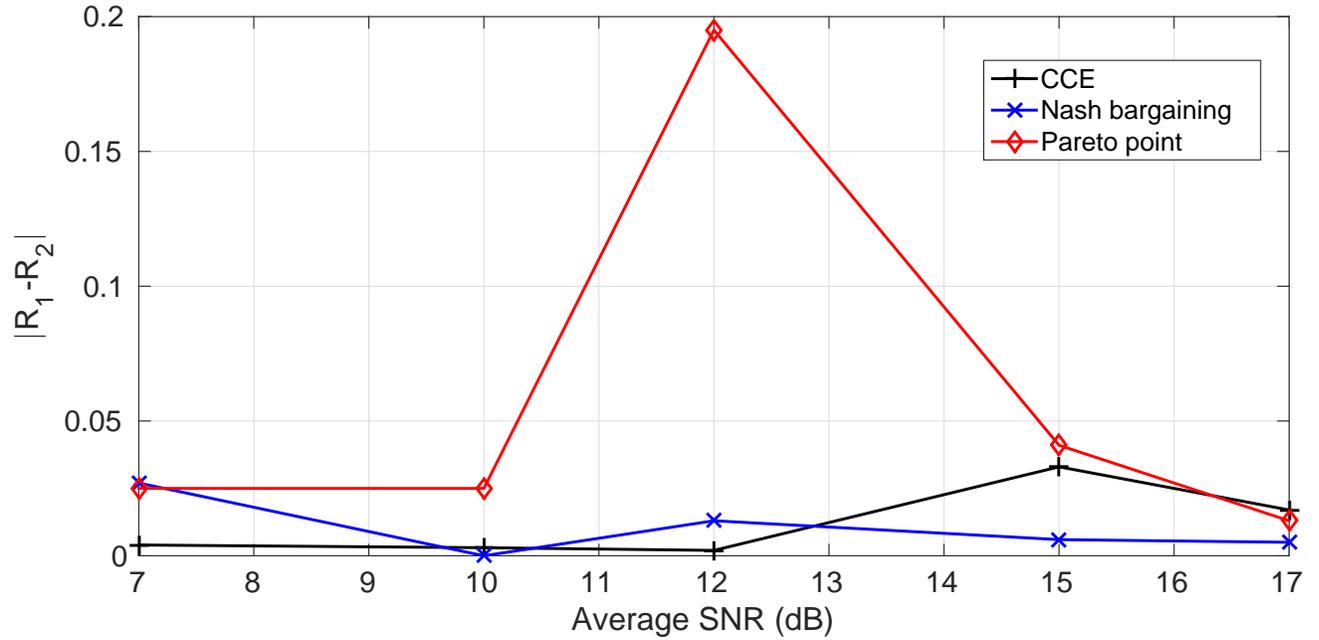}
               \caption{Comparing fairness of CCE, PP and NBS for F-MAC-WT, with no CSI of Eve (Fixed transmission rate)}
               \label{FMACWTFRFAIRNESSNOCSI}
               \vspace{-0.0 cm}
               \end{figure}
      

\section{Conclusions}
In this paper a $K$-user fading multiple access channel with and without security constraints  is studied. First we consider a F-MAC without the security constraints. Under the assumption of individual CSI of users, we propose the problem of power allocation as a stochastic game when the receiver sends an ACK or a NACK depending on whether it was able to decode the message or not. We have used Multiplicative weight no-regret algorithm to obtain a Coarse Correlated Equilibrium (CCE). Then we consider the case when the users can decode ACK/NACK of each other. In this scenario we provide an algorithm to maximize the weighted sum-utility of all the users and obtain a Pareto optimal point. PP is socially optimal but may be unfair to individual users. Next we consider the case where the users can cooperate with each other so as to disagree with the policy which will be unfair to individual user. We then obtain a Nash bargaining solution, which in addition to being Pareto optimal, is also fair to each user.
 \par
 Next we study a $K$-user fading multiple access wiretap Channel with CSI of Eve available to the users. We use the previous algorithms to obtain a CCE, PP and a NBS.
  Next we consider the case where each user does not know the CSI of Eve but only its distribution. In that case we use secrecy outage as the criterion for the receiver to send an ACK or a NACK. Here also we use the previous algorithms to obtain a CCE, PP or a NBS. Finally we show that our algorithms can be extended to the case where a user can transmit at different rates. At the end we provide a few examples to compute different solutions and compare them under different CSI scenarios.

\bibliographystyle{IEEEtran} 
\bibliography{single} 
\end{document}